# Highly Efficient Optical Add-Drop Filter With an Angle-Polished Fiber Coupler

Royce Dong, *Student Member*, *IEEE*, Zheng Fan, Jie Liao, Abraham Qavi, Gui-Lu Long, and Lan Yang, *Member*, *IEEE*

*Abstract*—Microbubble whispering-gallery resonators have shown great promise in fiber-optic communications because of their low confinement loss and hollow cores, which allow for facile stress-based tunability. Usually, the transmission spectrum of taper-coupled microbubbles contains closely spaced modes due to the relatively large radii and oblate geometry of microbubbles. In this letter, we develop an optical add-drop filter using a microbubble coupled to fiber taper and angle-polished fiber waveguides. Because of the extra degree of freedom in the angle of its polish surface, the angle-polished fiber can be used for the discriminatory excitation of certain radial-order modes in the optical microcavity, reducing the high modal density and enhancing add-drop selectivity. Our robust and tunable add-drop filter demonstrated a drop efficiency of 85.9% and quality factor of 2 x 10$^7$, corresponding to a linewidth of 9.68 MHz. As a proof of concept, the drop frequency was tuned using internal aerostatic pressure at a rate of 7.3 ± 0.2 GHz/bar with no diminishing effects on the add-drop filter performance.

## I. Introduction

Optical whispering gallery mode (WGM) microresonators [1], whose ultra-high quality factors (Q) and small modal volumes allow for the compact confinement of high optical power, have found use in many photonics applications such as lasing [2], [3], nonlinear optics [4], sensing [5], and fiber-optic communications [1], [6]. In the case of optical multiplexing, add-drop filters (ADFs) have been demonstrated with microtoroid [6], [7], microsphere [8], [9], microring [10], [11], photonic crystal [12], [13], and microbubble [14], [15] WGM resonators. Microbubbles are desirable for their narrow linewidth [16] and convenient stress-based tunability. The resonant wavelength can be tuned by elastically distorting the hollow microbubble structure, such as by applying internal aerostatic pressure [17], [18] or stretching the ends lengthwise with a piezo [14]. These methods are advantageous compared with conventional temperature-based tuning of ADFs, decreasing wavelength tuning response time and eliminating the need for heating elements.

Light is typically coupled into microbubbles with the fiber taper coupler [14]–[16]. The transmission spectrum of the taper-coupled microbubble usually displays a high modal density [16], [19] because the oblate profile and high modal volume of the microbubble support many nondegenerate axial modes. A high modal density diminishes add-drop selectivity and performance. Previous solutions to reducing the modal density include changing the taper's coupling position [20]–[22] or engineering a dampening element [19], [23], however these introduce additional coupling or scattering losses that broaden the linewidth and reduce the drop efficiency. In this letter, we demonstrate a microbubble-based ADF using fiber taper and angle-polished fiber (APF) couplers. The APF is essentially a prism coupler combined with a fiber waveguide [24], making it more robust than the taper and easily integrable into fiber-based photonic circuits without the need for collimating optics. We investigate the phase-matching requirements for the APF that enable the discriminatory excitation of WGMs compared with the taper, reducing the modal density in the transmission spectra. This will provide a new approach to selectively couple modes of interest to the dropping channel of ADFs.

## II. Principle

Fig. 1 depicts the configuration of the ADF composed of a microbubble resonator between a fiber taper and an APF. The WGM modes in the microcavity couple to the waveguide modes in the APF according to the phase-matching condition [24] $\cos \varphi = n_{microbubble}/n_{fiber}$ where $\varphi$ is the angle of the polished surface as defined in Fig. 1, $n_{microbubble}$ is the effective refractive index for circumferentially propagating WGMs in the resonator, and $n_{fiber}$ is the effective refractive index for waveguide modes in the APF. The polish angle $\varphi$ of the APF coupler can be chosen to differentially excite certain radial-order WGMs in the microcavity. This degree of freedom is not accessible in the fiber taper coupler.

The fiber taper acts as the bus waveguide and the APF acts as the add/drop waveguide in the ADF. The transmission coefficient T can then be expressed as [7]

$$T_{1\to 2} = \frac{P_{through}}{P_{in}} = \frac{4\Delta^2+(\kappa_0-\kappa_1+\kappa_2)^2}{4\Delta^2+(\kappa_0+\kappa_1+\kappa_2)^2} \quad (1)$$

This work was supported in part by the Army Research Office under grant W911NF1710189.

R. D., J. L., A. Q., and L. Y. are with the Department of Electrical & Systems Engineering, Washington University, St. Louis, MO 63130, USA (e-mail: roycedong@wustl.edu;  liaojie@wustl.edu;  abrahamqavi@wustl.edu; lyang25@wustl.edu).

Z. F. and G. L. are with the State Key Laboratory of Low-Dimensional Quantum Physics and Department of Physics, Tsinghua University, Beijing 100084, China (e-mail: zfphys@gmail.com; gllong@mail.tsinghua.edu.cn).

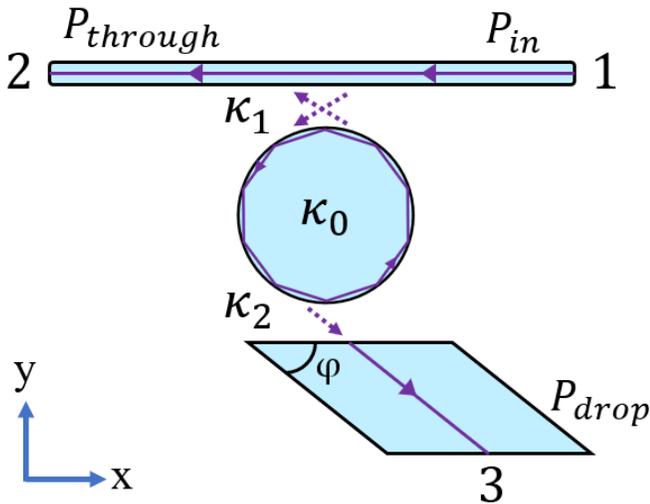

Fig. 1. Schematic of the add-drop filter with a microbubble resonator (equatorial cross-section), fiber taper, and angle-polished fiber. Optical power $P_{in}$ is inputted at port 1. Off-resonant light of power $P_{through}$ passes through the fiber taper at port 2. Resonant light of power $P_{drop}$ is dropped into the APF (with polish angle $\varphi$) to port 3. The total intrinsic resonator loss is $\kappa_0$, the microbubble-taper coupling loss is $\kappa_1$, and the microbubble-APF coupling loss is $\kappa_2$. Drawing is not to scale.

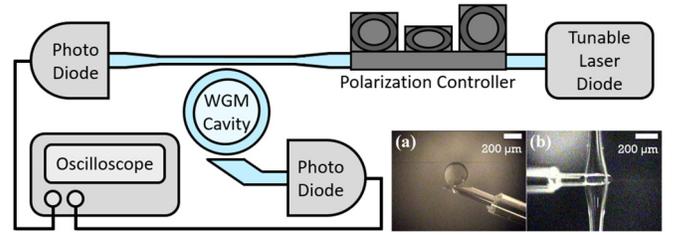

Fig. 2. Experimental setup for the characterization of the ADF. A tunable laser diode inputs light centered at 1550 nm with a scanning range of 30 GHz. The fiber-based polarization controller is used to optimize coupling. Light of off-resonant wavelengths is collected by the photodiode at the taper output port, and light of resonant wavelengths couples into the WGM microcavity and is collected by the photodiode at the APF output port. Photodiode signals are displayed by an oscilloscope. The inset shows optical images of the ADF from the top-view microscope (a) and from the sideview microscope (b).

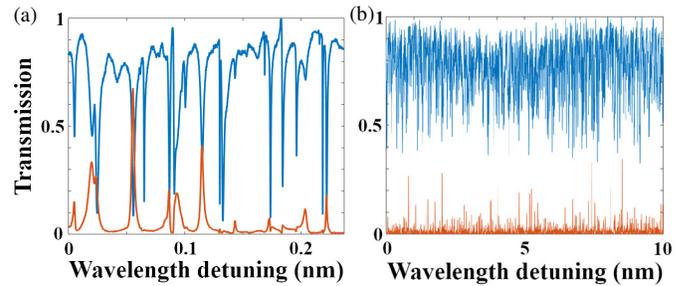

Fig. 3. Microbubble modal reduction with the APF. The top curve corresponds to the transmission spectrum through port 2 via the fiber taper waveguide, and the bottom corresponds to the spectrum through port 3 via the APF. (a) A typical ADF transmission spectrum over 0.24 nm (30 GHz). (b) A typical spectrum over 10 nm, which covers multiple free-spectral ranges (~2.0 nm) of the microbubble. The modal density of the taper transmission is 707 modes/nm, and that of the APF transmission is 191 modes/nm.

where $P_{in}$ is the input power at port 1, $P_{through}$ is the collected power at port 2, $\kappa_0$ is the total intrinsic loss for the resonator (combined material, scattering, and radiation losses), $\kappa_1$ and $\kappa_2$ are the coupling losses between the microbubble and the two couplers, and $\Delta$ is the detuning between the laser and resonance frequencies. Transmission at resonance, when $\Delta = 0$, simplifies to

$$T_{1\to 2} = \frac{(\kappa_0 - \kappa_1 + \kappa_2)^2}{(\kappa_0 + \kappa_1 + \kappa_2)^2} \qquad (2)$$

and denotes the crosstalk of the ADF. Similarly, drop efficiency coefficient $D_{1\to 3}$ at resonance can be written as

$$D_{1\to 3} = \frac{P_{drop}}{P_{in}} = \frac{4\kappa_1 \kappa_2}{(\kappa_0 + \kappa_1 + \kappa_2)^2} \qquad (3)$$

where $P_{drop}$ is the power dropped at port 3 and $P_{in}$ is the power inputted at port 1. From Eq. (2), the critical coupling condition for zero crosstalk is satisfied when $\kappa_1 = \kappa_0 + \kappa_2$. Treating intrinsic resonator loss $\kappa_0$ as fixed while coupling is being tuned, the drop efficiency can be maximized when $\kappa_1 = \kappa_2$ and $\kappa_1, \kappa_2 \gg \kappa_0$. However, such conditions increase coupling losses, decrease loaded Q, and broaden the linewidth of the ADF. Thus, both Q and $D_{1\to 3}$ cannot be simultaneously maximized. Tuning of these parameters requires precise control of $\kappa_0$, $\kappa_1$, and $\kappa_2$.

III. EXPERIMENT

We fabricated microbubbles with fused silica capillary tubing (Polymicro Technologies, 150 µm OD, 75 µm ID) using the fuse-and-blow technique [17]. Our microbubble resonator had an outer diameter of 256 µm and an equatorial wall thickness of 19 µm. The fiber taper was fabricated by heating a single-mode optical fiber with a hydrogen flame while pulling it apart axially [22]. The taper had a diameter of ~2 µm at its waist. The APF was fabricated by cleaving a single-mode optical fiber at a right angle and polishing it with 1-µm diamond lapping paper until the desired angle was achieved. Our APF had a polish angle of 19.9°, which we predicted would optimally excite first-radial-order modes in our fabricated microbubble based on its dimensions and elongated profile.

The fiber taper was positioned in the x-direction defined in Fig. 1, tangent to the equator of the spherical microcavity. The microbubble was positioned on an xyz-stage with its polar axis aligned vertically in the z-direction. The xyz-stage was used to fine-tune the coupling gap between the microbubble and the fiber taper. The APF was mounted on another xyz-stage, which was used to fine-tune the coupling gap between the APF and the microbubble. Fig. 2 outlines the experimental setup used to characterize the microbubble ADF. A tunable laser centered at 1550 nm was inputted at port 1 with a scanning range of 30 GHz, which coincides with the typical telecommunications wavelength band. The transmission spectrum through the fiber taper was collected at port 2. Light of resonant wavelength coupled into the resonator and was subsequently collected by the APF at port 3. WGM resonances are reflected by dips in the port 2 transmission and peaks in the port 3 transmission.

Fig. 3(a) shows a typical transmission spectrum for the add-drop configuration over 0.24 nm. The top curve corresponds to the transmission spectrum propagating through the fiber taper

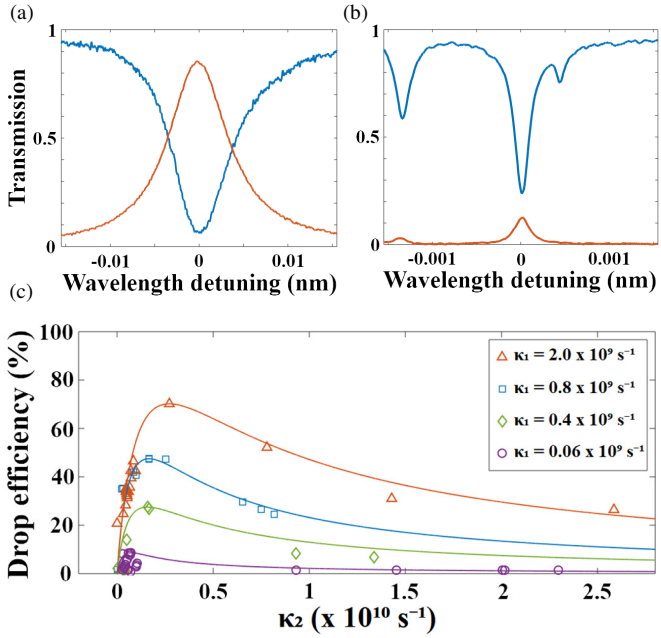

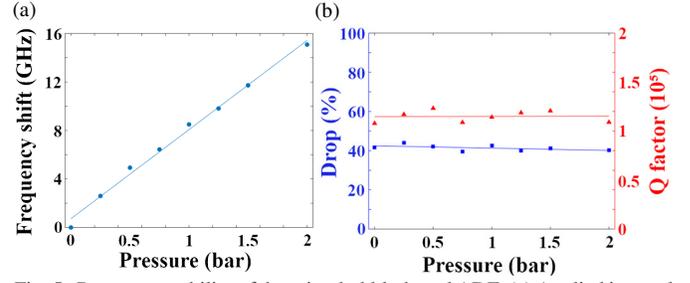

Fig. 4. Tuning of the ADF performance. (a) ADF transmission spectrum with drop efficiency D of 85.9% and Q factor of 2 x 10$^5$. (b) ADF spectrum with D of 16.6% and Q of 2 x 10$^7$. (c) Loading curves with D as a function of $\kappa_2$. The intrinsic resonator loss $\kappa_0$ is fixed at 1.0 x 10$^9$ s$^{-1}$, and the first coupling loss $\kappa_1$ varies from 0.06 x 10$^9$ s$^{-1}$ to 2.0 x 10$^9$ s$^{-1}$. The plots are curve fitted with $\kappa_0$, $\kappa_1$, and $\kappa_2$, and Eq. (3).

Fig. 5. Pressure tunability of the microbubble-based ADF. (a) Applied internal aerostatic pressure shifted the drop frequency with a sensitivity of 7.3 ± 0.2 GHz/bar. (b) Drop efficiency (squares) and Q factor (triangles) remained constant at 41.7 ± 2.2% and 1.10 ± 0.08 x 10$^5$ respectively, over the 2 bars of applied pressure.

(Port 2), and the bottom curve corresponds to the spectrum collected by the APF (Port 3). There is clearly not a one-to-one correspondence between the two spectra. The waveguide modes in the fiber taper couple to many WGMs in the cavity, yielding many dips. Of those, only a fraction couple out of the microcavity to the APF waveguide modes, yielding fewer peaks with varying intensities. Fig. 3(b) shows a typical transmission spectrum over 10 nm with coupling conditions different from those for Fig. 3(a). The free spectral range for the microbubble is estimated to be ~2.0 nm. The modal density for each spectrum is computed by counting the peaks/dips that exceeded the baseline noise level when there is no coupling to the microbubble. The modal density of the taper spectrum is measured as 707 modes/nm and that of the APF spectrum is 191 modes/nm, corresponding to a reduction in modal density of ~73%. The significant difference in WGM excitation is attributed to the extra degree of freedom in the polish angle $\varphi$ for the APF, which allows for the selective excitation of certain groups of WGMs different from those excited by the fiber taper.

The loading curves were obtained for the same microbubble, fiber taper, and APF configuration. Starting at no coupling, the microbubble was moved closer to the fiber taper in steps of 0.05 µm until physical contact was made. The intrinsic resonator loss $\kappa_0$ for the microbubble at the fixed z-position was found from the Lorentzian line shape as the coupling evolved from the deep undercoupling to deep overcoupling regimes. The coupling gap between the fiber taper and the resonator was then kept in the deep overcoupling regime at the fixed $\kappa_0$, and the APF was introduced with the other xyz-stage.

Fig. 4(a) highlights the spectrum with the high drop efficiency of 85.9% and Q of 2 x 10$^5$. Fig. 4(b) highlights the spectrum with drop efficiency of 16.6% and high Q of 2 x 10$^7$, which corresponds to a linewidth of 9.68 MHz (0.0775 pm). The drop efficiency D is calculated as the ratio of the collected power at port 3 (height of the peak) over the input power at port 1 (height of the dip). Fig. 4(c) depicts the loading curves for the ADF configuration. We recorded drop efficiency D while varying $\kappa_2$ (the coupling gap between the APF and microbubble) using the xyz-stage in steps of 0.05 µm. Moving the APF closer to/farther from the microbubble increased/decreased $\kappa_2$, while $\kappa_0$ and $\kappa_1$—the coupling configuration (xyz-positions) between the microbubble and fiber taper—were held constant. The data were collected for different $\kappa_1$ values by changing the x-position of the fiber taper relative to the microbubble and APF, thus altering the taper thickness at the coupling position. Executing the add-drop analysis in this way identifies the values for $\kappa_0$, $\kappa_1$, and $\kappa_2$, which are used for curve fitting the plots with Eq. (3). Intrinsic resonator loss $\kappa_0$ is measured to be 1.0 x 10$^9$ s$^{-1}$, and coupling loss $\kappa_1$ varies from 0.06 x 10$^9$ s$^{-1}$ to 2.0 x 10$^9$ s$^{-1}$. These results match with the theory and suggest that the tradeoff between drop efficiency, crosstalk, and quality factor can be easily optimized in the APF-taper system by tuning the two coupling gaps. The presented ADF possessed high stability during the measurement of the loading curves, and the positions of the microresonator and two couplers can be tuned independently of each other to achieve precise and reproducible configurations.

As a proof of concept, we demonstrate the stress-based tunability of the ADF. Internal aerostatic pressure can expand the microbubble structure and alter the optical path of the WGMs, thus red-shifting their resonant wavelengths [17]. The pressure tuning sensitivity is proportional to the microbubble diameter and inversely proportional to its wall thickness. The microbubble used had a major diameter of ~592 µm and wall thickness of ~6.5 µm. Based on these dimensions and the mechanical properties of fused silica, the theoretical pressure sensitivity was calculated to be 7.1 GHz/bar. One end of the microbubble was sealed during fabrication while the other end was connected to a closed pneumatic system, with a two-valve pressure gauge and piston controlling the internal gas pressure. The fiber taper and APF maintained physical contact with the microresonator, ensuring that the coupling gaps did not change during the expansion of the microbubble. The taper and APF transmission spectra were recorded as pressure was increased from 0 to 2 bars. Fig. 5(a) depicts the mode frequency shift as a function of pressure. The linear fitting reveals a pressure sensitivity of 7.3 ± 0.2 GHz/bar, which agrees with the theoretical value. Fig. 5(b) shows that the drop efficiency remained constant at 41.7 ± 2.2% and Q remained constant at $1.10 \pm 0.08 \times 10^5$ over the pressure range. There is thus no dependency of drop efficiency or linewidth on the applied pressure or drop frequency. The drop frequency can be tuned over tens of GHz without affecting the ADF performance.

## IV. CONCLUSION

In conclusion, we have demonstrated a highly efficient and tunable ADF using a microbubble resonator with an APF coupler. The hollow core of the microbubble lends it facile stress-based tunability of the drop frequency without diminishing the ADF performance. The polish angle φ of the APF coupler can be chosen to differentially excite certain radial-order WGMs in the microcavity. The APF coupler greatly reduces the high modal density of the taper-coupled spectrum by ~73%, thereby enhancing the add-drop selectivity. We report a drop efficiency as high as >85% and Q as high as $2 \times 10^7$. Future applications can use the capillary technology of hollow WGM resonators to support various fluids and gases for different material advantages, such as higher thermal stability, to improve the ADF. The highly efficient and robust APF can be used to couple to different resonator geometries, e.g. microtoroids and microspheres, for efficient and discriminatory excitation of WGMs.

ACKNOWLEDGMENT

R.D. Author wants to thank Yihang Li, Weijian Chen, Guangming Zhao, and Xiangyi Xu for their technical assistance on this project.

## REFERENCES

[1] K. J. Vahala, "Optical microcavities," *Nature*, vol. 424, no. 6950, pp. 839–846, 2003.
[2] J. Zhang et al., "A phonon laser operating ats an exceptional point," *Nat. Photonics*, vol. 12, no. 8, pp. 479–484, 2018.
[3] Y. Yang et al., "Tunable erbium-doped microbubble laser fabricated by sol-gel coating," *Opt. Express*, vol. 25, no. 2, pp. 428–432, 2017.
[4] Y. Li, X. Jiang, G. Zhao, and L. Yang, "Whispering gallery mode microresonator for nonlinear optics," *arXiv:1809.04878*, 2018.
[5] X. Jiang, A. J. Qavi, S. H. Huang, and L. Yang, "Whispering gallery microsensors: a review," *arXiv:1805.00062*, 2018.
[6] F. Monifi, S. Kaya Özdemir, and L. Yang, "Tunable add-drop filter using an active whispering gallery mode microcavity," *Appl. Phys. Lett.*, vol. 103, no. 181103, 2013.
[7] F. Monifi, J. Friedlein, S. K. Özdemir, and L. Yang, "A robust and tunable add-drop filter using whispering gallery mode microtoroid resonator," *J. Light. Technol.*, vol. 30, no. 21, pp. 3306–3315, 2012.
[8] M. Cai, G. Hunziker, and K. Vahala, "Fiber-Optic Add–Drop Device Based on a Silica Microsphere-Whispering Gallery Mode System," *IEEE Photonics Technol. Lett*., vol. 11, no. 6, pp. 686–687, 1999.
[9] P. Wang et al., "Packaged Optical Add-Drop Filter Based on an Optical Microfiber Coupler and a Microsphere," *IEEE Photonics Technol. Lett*., vol. 28, no. 20, pp. 2277–2280, 2016.
[10] T. Barwicz et al., "Microring-resonator-based add-drop filters in SiN: fabrication and analysis," *Opt. Express*, vol. 12, no. 7, p. 1437, 2004.
[11] X. Wang et al., "Wide-range and fast thermally-tunable silicon photonic microring resonators using the junction field effect," *Opt. Express*, vol. 24, no. 20, p. 23081, 2016.
[12] Z. Qiang and W. Zhou, "Optical add-drop filters based on photonic crystal ring resonators," *Opt. Express*, vol. 15, no. 4, pp. 1823–1831, 2007.
[13] M. Niyazi, A. Amirkhani, and M. R. Mosavi, "Investigation and simulation of a two-channel drop filter with tunable double optical resonators," *J. Supercond. Nov. Magn.*, vol. 27, no. 3, pp. 827–834, 2014.
[14] Z. H. Zhou, Y. Chen, Z. Shen, C. L. Zou, G. C. Guo, and C. H. Dong, "Tunable Add-Drop Filter with Hollow Bottlelike Microresonators," *IEEE Photonics J.*, vol. 10, no. 2, 2018.
[15] Y. Yin, Y. Niu, L. Dai, and M. Ding, "Cascaded Microbottle Resonator and Its Application in Add–Drop Filter," *IEEE Photonics J.*, vol. 10, no. 4, 2018.
[16] M. Pöllinger, D. O'Shea, F. Warken, and A. Rauschenbeutel, "Ultrahigh-Q Tunable Whispering-Gallery-Mode Microresonator," *Phys. Rev. Lett*., vol. 103, no. 5, pp. 11916–11925, 2009.
[17] R. Henze, T. Seifert, J. Ward, and O. Benson, "Tuning whispering gallery modes using internal aerostatic pressure," *Opt. Lett*., vol. 36, no. 23, pp. 4536–4538, 2011.
[18] Y. Yang, S. Saurabh, J. M. Ward, and S. Nic Chormaic, "High-Q, ultrathin-walled microbubble resonator for aerostatic pressure sensing," *Opt. Express*, vol. 24, no. 1, p. 294, 2016.
[19] J. Liao, X. Wu, L. Liu, and L. Xu, "Fano resonance and improved sensing performance in a spectral-simplified optofluidic micro-bubble resonator by introducing selective modal losses," *Opt. Express*, vol. 24, no. 8, p. 8574, 2016.
[20] G. S. Murugan, J. S. Wilkinson, and M. N. Zervas, "Optical excitation and probing of whispering gallery modes in bottle microresonators: potential for all-fiber add–drop filters," *Opt. Lett*., vol. 35, no. 11, pp. 1893–1895, 2010.
[21] X. Jin, Y. Dong, and K. Wang, "Selective excitation of axial modes in a high-Q microcylindrical resonator for controlled and robust coupling," *Appl. Opt*., vol. 54, no. 27, pp. 8100–8107, 2015.
[22] V. Kavungal, G. Farrell, Q. Wu, A. K. Mallik, and Y. Semenova, "A comprehensive experimental study of whispering gallery modes in a cylindrical microresonator excited by a tilted fiber taper," *Microw. Opt. Technol. Lett.,* vol. 60, no. 6, pp. 1495–1504, 2018.
[23] A. A. Savchenkov, A. B. Matsko, D. Strekalov, V. S. Ilchenko, and L. Maleki, "Mode filtering in optical whispering gallery resonators," *Electron. Lett*., vol. 41, no. 8, 2005.
[24] V. S. Ilchenko, X. S. Yao, and L. Maleki, "Pigtailing the high-Q microsphere cavity: a simple fiber coupler for optical whispering-gallery modes," *Opt. Lett*., vol. 24, no. 11, pp. 723–725, 1999.